\documentclass[preprint,showkeys,superscriptaddress,amsmath,amssymb]{revtex4-2}

\usepackage{times}
\usepackage{graphicx}
\usepackage[version=3]{mhchem}
\usepackage{upgreek}
\usepackage{multirow}

\renewcommand{\figurename}{Figure}

\begin{document}

\begin{abstract}
  Metasurfaces enable precise control over the properties of light and
  hold promise for commercial applications. However, fabricating
  visible metasurfaces suitable for high-volume production is
  challenging and requires scalable processes. Nanoimprint lithography
  is a cost-effective and high-throughput technique that can meet this
  scalability requirement. This work presents a mask-templating
  nanoimprint lithography process for fabricating metasurfaces with
  varying fill factors and negligible wavefront aberrations using composite stamps. As a proof-of-concept, a 6 mm diameter  metalens formed of silicon nitride nano-posts with a numerical
  aperture of 0.2 that operates at 550 nm is demonstrated. The
  nanoimprinted metalens achieves a peak focusing efficiency of (81 $\pm$
  1)\%, comparable to the control metalens made with electron beam
  lithography with a focusing efficiency of (89 $\pm$ 1)\%. Spatially
  resolved deflection efficiency and wavefront data, which informs
  design and process optimization, is also presented. These results
  highlight nanoimprint lithography as a cost-effective, scalable
  method for visible metasurface fabrication that has potential for
  widespread adoption in consumer electronics and imaging systems.
\end{abstract}

\title{Visible metalenses with high focusing efficiency fabricated
  using nanoimprint lithography}

\newcommand{\UMassECE}{%
 \affiliation{%
  Department of Electrical and Computer Engineering,
   University of Massachusetts Amherst, Amherst, MA 01003, USA
 }
}

\newcommand{\UMassPSE}{%
 \affiliation{%
  Department of Polymer Science and Engineering,\\
  University of Massachusetts Amherst, Amherst, MA 01003, USA
 }
}

\author{Andrew McClung}
\UMassECE{}
\author{Mahsa Torfeh}
\UMassECE{}
\author{Vincent J. Einck}
\UMassPSE{}
\author{James J. Watkins}
\UMassPSE{}
\author{Amir Arbabi}
\email{arbabi@umass.edu}
\UMassECE{}
\maketitle

\section*{Introduction}
Metasurfaces are optical elements created by arranging subwavelength
scatterers (known as `meta-atoms') on a surface. Metasurfaces have
several desirable features: They can control the wavefront,
polarization, and amplitude of light with subwavelength
resolution~\cite{kamali2018review}. Compared to traditional refractive
optical elements, metasurfaces are significantly lighter and their
planar form factor simplifies integration into compact optical
systems. Furthermore, their design allows for cascading, enabling the
realization of optical
metasystems~\cite{arbabi2016miniature,groever2017meta,faraji2018compact,kwon2020single,mcclung2020snapshot,arbabi2023advances,neshev2023enabling}. These
advantages have sparked interest in the utilization of metasurfaces
for commercial applications~\cite{metalenzpr2021,niltpr2021}. However,
widespread adoption will require metasurface designs and fabrication
processes suitable for high-volume production.

For metasurfaces to perform well, the dimensions and positions of
their meta-atoms must be precisely controlled. Subwavelength critical
dimensions and spacing are needed to achieve high efficiency, which
presents challenges in patterning visible and shorter wavelengths
metasurfaces. Commonly, metasurface patterns for academic research are
defined using electron-beam lithography (EBL), a technique that meets
these precise positioning and resolution requirements. Unfortunately,
EBL has low throughput and is unsuitable for mass production. While a
fabrication process for metasurface lenses (metalenses) using
deep-ultraviolet photolithography, which is a scalable patterning
approach, has been demonstrated~\cite{park2019all}, the metalens
pattern in that study was constrained to minimum feature sizes equal
to the wavelength used in lithography (248 nm), limiting both
numerical aperture (NA) and focusing efficiency. Smaller features can
be achieved by using shorter wavelength illumination, multiple
patterning, or other advanced photolithographic techniques. However,
the use of these technologies increases the cost of manufacture
substantially~\cite{irds2022}.

Nanoimprint lithography (NIL) is a promising alternative to
photolithography for applications where costs of nanoscale patterning
dominate~\cite{irds2022}. NIL is a low-cost, high-throughput
manufacturing method in which a stamp with nanoscale features
mechanically deforms an imprint material to create a desired
pattern~\cite{schift2008nanoimprint}. While it is not yet competitive
with photolithography for semiconductor manufacture, current NIL
technology is viable for devices with less stringent demands in terms
of defect density, overlay accuracy, and throughput. This makes it an
attractive manufacturing method for optical
elements~\cite{barcelo2016nanoimprint} like
polarizers~\cite{yu2000reflective,ahn2005fabrication}, microring
resonators~\cite{chao2002polymer}, waveguides~\cite{koo2011liquid} and
photonic crystals~\cite{ji2010uv,beaulieu2014large}.

NIL is also a promising candidate for scalable metasurface
fabrication, as evidenced by several recent demonstrations. NIL
metasurface fabrication processes fall into two main categories:
direct
imprint~\cite{yoon2020single,einck2021scalable,miyata2022scalable,jung2023refractive}
and mask
templating~\cite{lee2018metasurface,dirdal2020towards,quaade2022highly}. In direct imprint,
the stamp shapes an ink that is subsequently cured to form the
metasurface.  This approach requires very few processing steps, which
minimizes both fabrication time and cost.  However, only a few
materials can be made into curable inks suitable for metasurfaces.

In mask templating, the stamp shapes a temporary layer that serves as
an etch or lift-off mask.  Although it requires more processing steps
than direct imprint, mask templating offers the advantage of using
material platforms proven in EBL workflows, as all other processing
steps (e.g., material deposition or dry etching) remain the
same. 

Here, we present a mask-templating NIL process using tri-layer composite stamps~\cite{odom2002} for patterning
metasurfaces with spatially varying fill factors that operate in the
visible wavelength range. The ability to pattern structures with
varying fill factors enables the realization of
polarization-insensitive metasurfaces operating based on propagation
phase, in contrast to geometric phase metasurfaces that are limited to
circularly polarized
light~\cite{yoon2020single,miyata2022scalable,lee2018metasurface,dirdal2020towards}. We
evaluate the fabrication process by designing and patterning a
polarization-insensitive silicon nitride metalens with a 6~mm diameter
and an NA of 0.2 at a design wavelength of 550 nm. Silicon
  nitride is an established material platform for visible wavelength
  metasurfaces~\cite{zhan2016low,colburn2018broadband,fan2018silicon}. The metalens produced by our process exhibits a nearly
  diffraction-limited spot and a focusing efficiency of (81 $\pm$ 1)\%. A   control metalens made using EBL exhibited a focusing efficiency of
  (89 $\pm$ 1)\%, indicating that our NIL metalens was fabricated with   high fidelity. Furthermore, to provide insights into the factors
contributing to the slightly lower focusing efficiency on NIL metalenses, we present spatially resolved deflection efficiency by
local probing and wavefront data obtained through a stable holographic
technique.
\section*{Results and discussion}
\begin{figure}
  \centering
  \includegraphics[width=\textwidth]{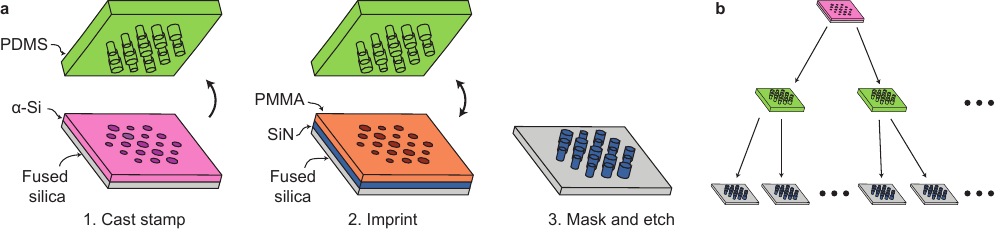}
  \caption{ Schematic of scalable NIL fabrication process. (a) A PDMS
    elastomer stamp is cast from a master mold comprising holes in
    thin film amorphous silicon on fused silica. The stamp is used to
    imprint a pattern in PMMA resist. The pattern is transferred to a
    silicon nitride layer using a hard mask and dry etch process. (b)
    A single master mold is used to make many stamps, each of which is
    used to make many metasurfaces.}
  \label{fig:concept}
\end{figure}
Our NIL patterning process, illustrated schematically in
Figure~\ref{fig:concept}a, involves using a hybrid silicone elastomer
stamp that is cast on a master mold to imprint a resist on a metalens
substrate. The thermal NIL process that we employ is a well-studied process with proven repeatability and reliability~\cite{chou1996}. To make metasurfaces, we evaporate an etch mask into the
resist, lift it off, and use the etch mask to transfer the pattern
into the substrate. Through this approach, a single master mold can be
used to make several stamps, each of which can pattern several wafers
by imprinting, enabling high-volume replication of a pattern
(Figure~\ref{fig:concept}b)~\cite{schmitt2012life,Tucher2017}.

\begin{figure}
  \centering
  \includegraphics[width=\textwidth]{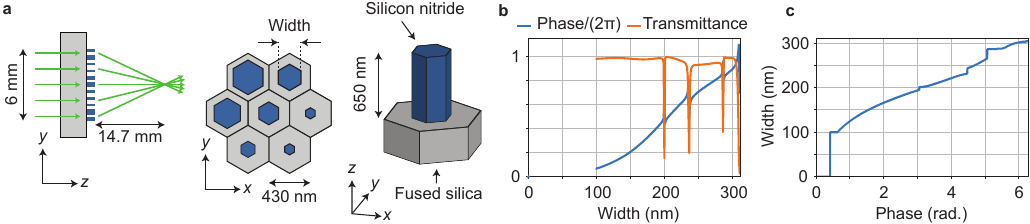}
  \caption{Metalens design. (a) A metalens with a diameter of 6 mm and
    a focal length of 14.7mm (NA of 0.2) is designed. The metalens is
    formed of hexagonal silicon nitride nano-posts arranged on a
    triangular lattice and resting on a fused silica substrate. (b)
    Phase and transmittance as a function of nano-post width. (c) The
    design curve that maps the desired phase to the nano-post width.}
  \label{fig:design}
\end{figure}

To showcase the process and assess the performance of the fabricated
metasurfaces, we opted to design, fabricate, and characterize
metalenses with a diameter of 6 mm and NA of 0.2, resulting in a focal
length of 14.7 mm, for operation at a wavelength of 550 nm
(Figure~\ref{fig:design}a). Metalenses that are corrected for spherical aberrations
have practical applications and serve as benchmarks for evaluating the
performance of different metasurface design and fabrication
techniques. The choice of a large aperture was deliberate, as the
magnitude of aberrations scale with the diameter of the aperture. We
anticipated that the NIL process might introduce more substantial
aberrations than EBL due to potential errors in the placement of the
zone boundaries caused by stamp stretching and deformation during the
stamp fabrication and the imprinting process. Therefore, a large
aperture metalens, which is more sensitive to such aberrations, is
suitable for evaluating the performance of the NIL process.

As shown in {Figure ~\ref{fig:design}}a, the metalens is composed of
silicon nitride nano-posts that are 650 nm in height. These nano-posts
are positioned on a fused silica substrate and have hexagonal
cross-sections. They are arranged on a triangular lattice with a
lattice constant of 430 nm, and their widths vary between 100 nm and
310 nm.  The minimum nano-post width and gap of 100 nm and 120 nm were not imposed by the NIL process but were selected to facilitate potential wafer-scale fabrication of the master mold using a cost-effective process such as  immersion deep ultraviolet (DUV) lithography. Figure~\ref{fig:design}b shows the simulated transmittance and phase characteristics for periodic nano-posts arrays as a function of
their width. To determine the design parameters and estimate the metalens focusing efficiency, we employed the grating averaging technique~\cite{arbabi2020increasing}. This technique allowed us to establish a design, depicted in Figure~\ref{fig:design}c, that correlates a desired phase with the width of the nano-post. We obtained a simulated focusing efficiency of 90.2\% for the metalens shown in Figure~\ref{fig:design}, which has an NA of 0.2. We also designed and simulated silicon nitride metalenses with larger NAs while limiting the minimum nano-post widths to 100 nm and the minimum gaps between them to 120 nm. The results are presented in Figure S1 and indicate that a focusing efficiency of more than 75\% can be achieved for metalenses having an NA smaller than 0.7.  

To create the metalens pattern, we fabricated a master mold by EBL and
dry etching holes in a 200-nm-thick film of amorphous silicon
deposited on a fused silica substrate {(see Supporting Information for
  details)}. In the fluorine plasma chemistry we use, the etch rate of
amorphous silicon is substantially higher than that of fused silica,
and the use of amorphous silicon on fused silica substrate ensures
that all the holes are etched to nearly the same depth. After
fabrication of the master mold, we fluorinated it to reduce adhesion
during stamp fabrication.

\begin{figure}
  \centering
  \includegraphics[]{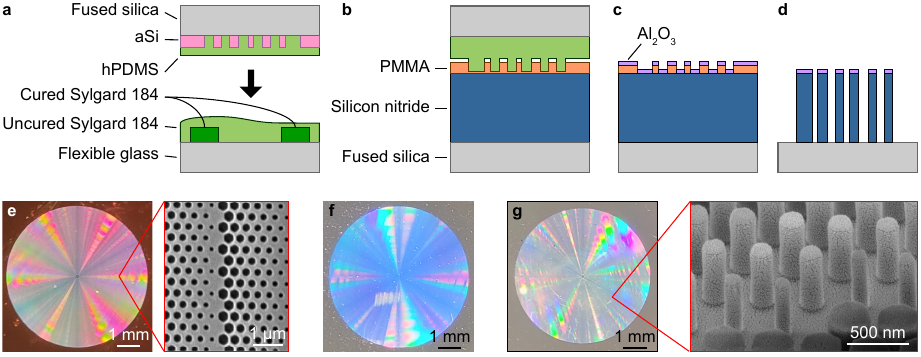}
  \caption{Fabrication process of NIL metalenses. (a) Stamp assembly:
    hPDMS is spin-coated on the master mold, partially cured, and then
    adhered to a flexible glass substrate by a commercial PDMS
    (Sylgard 184). Spacers set the thickness of the stamp. (b) Thermal
    NIL: The PDMS stamp is used to imprint features in PMMA. (c) Hard
    mask deposition: an Al2O3 mask is evaporated into the resist
    template and lifted off. (d) Etching: the hard mask is used to
    etch metalens patterns into silicon nitride. (e) Photograph and
    SEM image of the master mold, (f) a photograph of the stamp, and
    (g) a photograph and an SEM image of an imprinted metalens. The roughness
    observed in the SEM image is due to the gold coating that was done
    to mitigate charging during imaging.}
  \label{fig:fabrication}
\end{figure}

To make a stamp, we spun a hard PDMS (hPDMS) film onto the master mold
and cured it by baking it in an oven. We then placed the hPDMS side of
the mold in an uncured puddle of commercial PDMS that was resting on a
flexible glass backing, as illustrated in
Figure~\ref{fig:fabrication}a. Cured PDMS spacers with a thickness of
140 $\mu$m were employed to maintain a fixed distance between the master
mold and the glass backing. The use of bilayer PDMS with optimized thicknesses and the flexible yet inelastic glass backing ensures that the stamp does not stretch or deform during the imprint process because stamp deformation can lead to substantial metasurface wavefront aberrations. The entire assembly was then baked to cure
the PDMS, and the mold was released to yield a stamp comprising
nano-posts that are approximately 200 nm in height. Because the stamp
is used to template a mask, it does not need to have the same feature
height as the final device: all features in the stamp have an aspect
ratio of 2 or smaller. The small aspect ratio of the stamp features enables patterning with a resolution down to 10 nm~\cite{Pandey2019}.

To fabricate the metalenses, we first spun a layer of polymethyl
methacrylate (PMMA) on top of a silicon nitride film that was
deposited on a fused silica substrate. We then used a commercial
thermal NIL tool to imprint the metalens patterns into the PMMA
layer. The imprinting process is depicted in
Figure~\ref{fig:fabrication}b. To remove the thin residual layer of PMMA
at the bottoms of the features, we etched the imprint in oxygen
plasma. The optical and scanning electron microscope (SEM) images of
the imprinted PMMA displayed in Figure~S2 in the Supporting Information demonstrate
that the patterns were transferred with high fidelity and minimal
defects. Following the successful imprinting of the patterns, we
evaporated an aluminum oxide etch mask into the PMMA
(Figure~\ref{fig:fabrication}c) and lifted it off. Finally, we etched
the metalens patterns into the silicon nitride layer, resulting in the
fabrication of metalenses (Figure~\ref{fig:fabrication}d). Photographs
and SEM images of the master mold, a stamp and an imprinted metalens
are shown in Figure~\ref{fig:fabrication}e–g. For performance
comparison, we also produced a control metalens with the same design
as the NIL metalens but using EBL. In addition, the experimental results for a second NIL metalens that was fabricated using the same NIL process but with a sub-optimal silicon nitride etch process are presented in Figure S3. A more comprehensive description of
the fabrication processes for the master mold, stamps, metalenses, and
the EBL metalens can be found in the Supporting Information.
 
We characterized the performance of the NIL metalens by measuring its
focal spot and focusing efficiency. The focal spot shape mostly
provides information about wavefront aberrations caused by potential
errors in the zone boundary locations (i.e., positions of the
nano-posts) that could arise from the deformation of the stamp during
the imprinting process. Aberrations of this nature are typically not
observed in metalenses patterned using EBL or photolithography due to
their high positioning
accuracy~\cite{fischer199310,wallraff1999lithographic}. On the other
hand, the focusing efficiency, which represents the proportion of
incident power that is focused by the metalens, is used to evaluate
the accuracy of the nanostructure pattern generated through the NIL
process. The focusing efficiency is sensitive to the shapes and
dimensions of the meta-atoms, allowing for an assessment of the
fidelity of the nanostructure pattern created using NIL.

\begin{figure}
  \centering
  \includegraphics[]{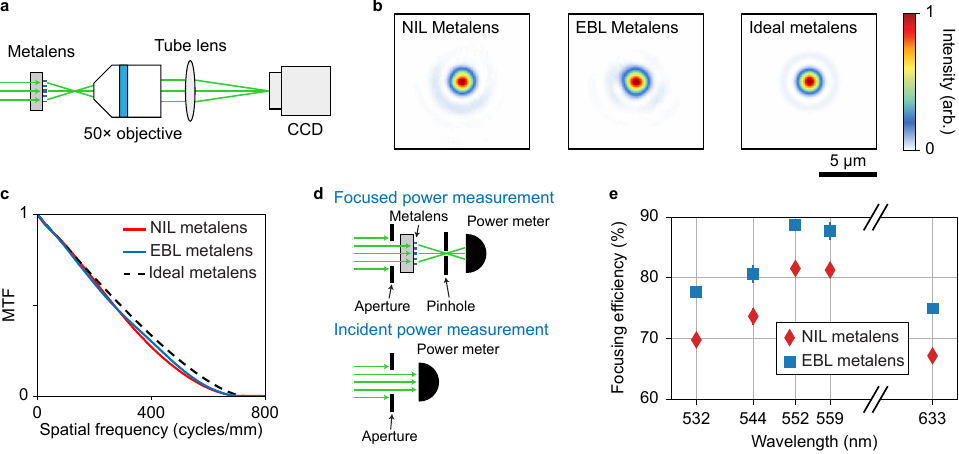}
  \caption{Metalens characterization. (a) Schematic of image relay
    used for measuring focal spots. (b) Focal spot images of NIL and
    EBL metalenses and that of an ideal metalens. (c) Modulation
    transfer functions of NIL, EBL, and ideal metalenses. (d)
    Schematic of measurement setups used for determining metalens
    focusing efficiency. (e) Measured focusing efficiency values for
    NIL and EBL metalenses at a few wavelengths between 532 and 633
    nm.}
  \label{fig:spot}
\end{figure}

We measured the focal spots of the NIL and EBL metalenses using a
custom-built microscope, as shown schematically in
Figure~\ref{fig:spot}a. We passed a laser beam with wavelength of 552 nm
through a beam expander and illuminated the metalenses, which had a
diameter of 6 mm and an NA of 0.2 at 550 nm. An objective lens and a
tube lens magnified and imaged the focal spots of metalenses onto a
camera {(see Supporting Information for details)}. Figure~\ref{fig:spot}b
presents the measured focal spot images of the NIL and EBL metalenses,
and the simulated focal spot of an ideal metalens with the same NA.  An ideal metalens is a local metalens that directs all the normally incident power to its focal spot without any spherical aberrations~\cite{ McClung2020}. The focal spots of ideal metalenses with an NA smaller than 0.8 can be well approximated with an Airy disk~\cite{ McClung2020}. 
We used the focal spot data in Figure~\ref{fig:spot}b to calculate the
modulation transfer function (MTF) of the metalenses. The MTFs of the
NIL and EBL metalenses, as well as the simulated MTF of a
diffraction-limited ideal metalens, are shown in
Figure~\ref{fig:spot}c. The similarity between the aberrations observed
in the focal spots of the NIL and EBL metalenses suggests that the
slight deviations from the diffraction limit may be due to the
nonuniform intensity distribution and wavefront aberrations of the
incident expanded beam. Nonetheless, the nearly diffraction-limited
response of the NIL metalens confirms the negligible error in the
position of the nano-posts. It is worth noting that when the process
parameters during the imprinting process were suboptimal, we observed
significant aberrations in the focal spots of the NIL
metalenses. Hence, optimizing the NIL process is essential for
achieving high-quality metasurfaces.

We also conducted measurements to determine the focusing efficiency of
the metalenses. The focusing efficiency is defined as the fraction of
power incident on the metalens that is focused by the metalens. In
Figure~\ref{fig:spot}d, we schematically depict the setup used for these
measurements. To quantify the optical power focused by the metalens,
we employed a 6-mm-diameter aperture positioned behind the metalens,
along with a 40-$\mu$m-diameter pinhole placed at the focal plane of the
metalens, and measured the optical power that passed through the
pinhole. The purpose of the aperture was to reduce the size of the
incident beam, matching it to the aperture size of the metalens. On
the other hand, the pinhole served to block the light that passed
through the metalens but did not converge to the desired focal spot,
including the zeroth and other transmission diffraction orders apart
from the first order. To ensure that the optical power passing through
the pinhole predominantly originated from the focused light and had
minimal contributions from unfocused light, we conducted additional
tests. Specifically, we obtained a high dynamic range image of the
intensity distribution at the focal plane of the NIL metalens and
compared it to the simulated intensity distribution of an ideal
metalens. {Detailed information about these procedures can be found in
  the Supporting Information}. Figure~S4 in the Supporting Information presents the resulting
intensity distributions, along with the encircled power percentage as
a function of the pinhole diameter. The simulated encircled power
percentage indicates that 98.6\% of the power focused by an ideal
metalens passes through the 40-$\mu$m-diameter pinhole.

We measured the incident power by removing the metalens and the
pinhole and measuring the optical power that passed through the
6-mm-diameter aperture. To determine the focusing efficiency, we
divided the power transmitted through the pinhole by the incident
power and accounted for corrections due to the Fresnel reflection at
the fused silica--air interface at the backside of the metalens
substrate (3.5\%) and the percentage of power focused by an ideal
metalens that is blocked by the pinhole
(1.4\%). Figure~\ref{fig:spot}e shows the measured values of the
focusing efficiency for NIL and EBL metalenses at wavelengths between
532 nm and 633 nm. The NIL metalens exhibits a peak focusing
efficiency of (81 $\pm$ 1)\% at the wavelength of 552 nm which is close to
the (89 $\pm$ 1)\% obtained for the EBL metalens at the same
wavelength. The focusing efficiency data are also presented in
Table S1 in the Supporting Information. The high focusing efficiency observed for the
NIL metalens validates the high accuracy of the nanostructure patterns
produced using the NIL technique. This outcome is consistent with the
SEM inspection results depicted in Figure~\ref{fig:fabrication}g,
affirming the suitability of NIL for fabricating high-quality visible
wavelength metasurfaces.

The focusing efficiency, which is a coherent average of efficiency
over the metalens aperture, does not capture the potential efficiency
variation across the aperture or provide insights into its underlying
causes. Identifying and understanding potential causes of efficiency
variation are crucial for maximizing the metalens efficiency by
adjusting the metalens design or fabrication parameters such as the
NIL process or etching parameters. To detect such variations and gain
insights into their potential causes, we conducted two additional
measurements: local deflection efficiency and high-resolution
holographic complex field measurements. We conducted deflection
efficiency measurements at various locations on the NIL metalens by
illuminating small regions of the aperture with a loosely focused beam
(ca.~50~$\mu$m diameter). The purpose of these measurements was to
determine the percentage of power directed toward the metalens' focal
point. Figure~S5 in the Supporting Information presents the results, which demonstrate
that the deflection efficiency rises as the distance from the center
of the metalens increases.

\begin{figure}
  \centering
  \includegraphics[]{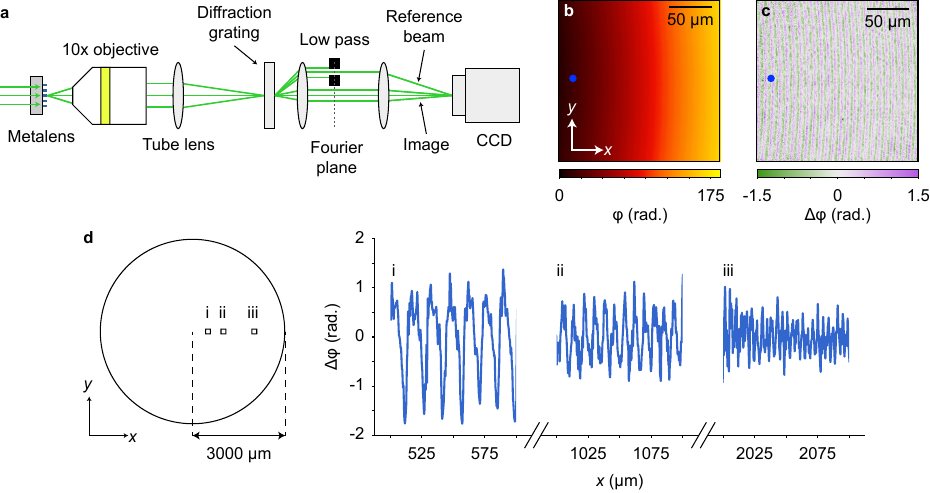}
  \caption{Holographic characterization. (a) Schematic of the optical
    setup used for holographic complex field measurement. (b) Measured
    phase profile over a small region on the surface of metalens. The
    blue dot indicates the location of (x, y) = (1000 $\mu$m, 0 $\mu$m)
    assuming the center of the metalens is at (x, y) = (0 $\mu$m, 0
    $\mu$m). (c) The Difference between the measured and ideal phase over
    the region shown in (b). (d) Left: Diagram of metalens showing
    regions of interest. Regions i, ii and iii lie at distances of 500
    $\mu$m, 1000 $\mu$m and 2000 $\mu$m from the metalens center. Right: Line cuts
    of the measured and ideal phase difference in regions i, ii, and
    iii.}
  \label{fig:holographic}
\end{figure}

The spatially resolved deflection efficiency measurement provides
valuable information and aligns with the focusing efficiency values
obtained using the pinhole. However, it does not reveal the underlying
causes of the lower efficiency observed near the center of the
metalens. To further investigate this issue, we employed a
high-resolution digital holography technique
\cite{edwards2012optically,popescu2006diffraction,
  bhaduri2014diffraction} to measure the complex-valued optical field
on the output aperture of the metalens. Figure~\ref{fig:holographic}a
illustrates the schematic of the holographic microscope used in the
experiment. A coherent illumination at 552 nm was projected onto a
diffraction grating, and the resulting image of the metalens surface
was captured after passing through a relay. By filtering the first
diffraction order of the grating in the Fourier plane of the relay, a
coherent reference beam was created, which interfered with the relayed
image. We captured images of the metalens surface and utilized a
Fourier technique to obtain a high-resolution map of the
complex-valued optical field on the metalens surface. {The holographic
  imaging setup and measurements are described in greater detail in
  the Supporting Information.}

The extracted phase of the field at a region approximately 1 mm from
the lens center is shown in Figure~\ref{fig:holographic}b, and the phase
error, which is the difference between the measured phase and the
desired phase profile, are shown in Figure~\ref{fig:holographic}c. Line
cuts showing the phase error at different distances from the metalens
center are presented in Figure~\ref{fig:holographic}d. Notably, larger
phase errors are noticeable near the center of the metalens, but they
decrease in magnitude as the radius increases, in line with the trend
observed in the local deflection efficiency results. Moreover, the
phase error indicates that the realized phase coverage over a Fresnel
zone exceeds $2\pi$. This could be attributed to factors such such as nano-post widths being larger than the design values. To further quantify the effect of such errors, we used the grating averaging technique~\cite{arbabi2020increasing}, and found the focusing efficiency of the metalens as a function of systematic errors in the nano-post height and width, and random errors in the nano-post widths. The results of this study are presented in Figure S6 and indicate that the focusing efficiency is more sensitive to errors in the nano-post widths than their height.

\begin{figure}
  \centering
  \includegraphics[]{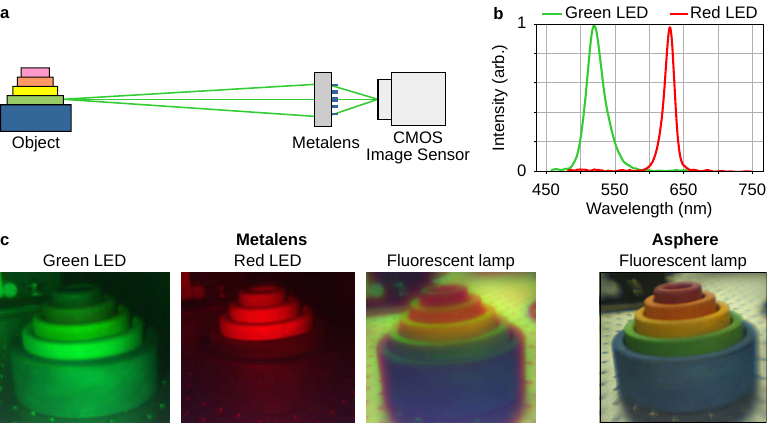}
  \caption{Imaging demonstration. (a) Schematic of the imaging
    setup. Light emanating from a distant object is focused by the
    metalens onto a color CMOS image sensor. (b) Spectra of narrowband
    green and red LED illumination sources used in imaging. (c)
    Comparison of photos taken with a 6-mm-diameter metalens under
    narrowband green, narrowband red, and white light illumination,
    and a refractive aspheric lens. The photos represent the raw data
    captured by the image sensor, and no processing has been performed
    on them.}
  \label{fig:imaging}
\end{figure}

The NIL metalens was primarily designed for optimal performance at a
wavelength of 550 nm. However, the focusing efficiency values depicted
in Figure~\ref{fig:spot}e indicate that it can also produce
high-contrast images at other wavelengths within the visible
spectrum. To demonstrate its imaging properties, we assembled a camera
using the metalens and an RGB image sensor, shown schematically in
Figure~\ref{fig:imaging}a: we illuminated the object with different
light sources—two narrowband LED sources, whose spectra are shown in
Figure~\ref{fig:imaging}b, and the fluorescent lamps installed in our
lab—placed the metalens camera assembly approximately 30 cm from the
object, and adjusted the distance between metalens and image sensor to
obtain the best focus. Figure~\ref{fig:imaging}c shows images captured
with our camera. Clear images are formed when the object is
illuminated with narrowband green and red light sources, and the image
formed under broadband illumination exhibits the chromatic aberration
expected for metalenses that employ phase wrapping. To provide a color
reference and for comparison, we imaged the object using the same
image sensor and an aspheric lens with similar NA
(Figure~\ref{fig:imaging}c, right). The varying focusing efficiency of
the NIL metalens across the visible spectrum can alter the apparent
color of objects within the scene. However, this effect can be readily
corrected by adjusting the white balance of the image, which is a
digital post-processing step that has not been performed on photos
shown in Figure~\ref{fig:imaging}c.

\section*{Conclusion and outlook}
The scalable fabrication of high-efficiency metasurfaces operating in
the visible wavelength range opens up opportunities for their
widespread adoption in consumer applications. Unlike some of the
previous NIL techniques developed for visible metasurfaces, the
approach described in this study offers distinct advantages by not
imposing limitations on the metasurface materials or their patterns,
including fill factors and shapes. Our methodology employs an
amorphous silicon on fused silica master mold, which  ensures
consistent height of the stamp features, along with a thin hard mask
that allows for the utilization of a thin PMMA layer. This
combination enables precise patterning of nanostructures with
spatially varying fill factors, achieving high fidelity in the
process. The technique's high resolution facilitated the patterning of
small features for visible metasurfaces, and it could potentially be
extended to pattern metasurfaces even at UV wavelengths.

At scale, this fabrication method could significantly reduce the cost
of producing visible-wavelength metalenses compared to equivalent
metalenses created using photolithography. Furthermore, the nano-post
meta-atoms utilized in this study enable the implementation of
metasurfaces that not only manipulate the phase of light but also
control its polarization. As a result, the NIL technique holds promise
for the cost-effective fabrication of such elements, expanding the
range of applications and functionalities achievable in metasurface
devices. In fact, while preparing this work for publication~\cite{mcclung2022nanoimprint}, we learned that a mask templating process for manufacturing niobium oxide metalenses, though with a lower efficiency and a smaller diameter, has been recently developed by a commercial company~\cite{williams2023metalenses}, further validating the advantages of the NIL technique for the manufacturing of metalenses.

In numerous applications, the efficiency of metasurfaces is of utmost
importance, making its accurate measurement crucial. The approach
discussed here, which involves direct focusing efficiency measurement
by measuring the optical power passing through a small-diameter
pinhole, combined with high dynamic range imaging for precise capture
of the transmitted light, can serve as the standard technique for
comparing and benchmarking different design and fabrication methods
for metalenses. It is worth noting that when measuring high numerical
aperture (NA) metalenses, a power meter equipped with an integrating
sphere may be required to ensure accurate measurement of the focused
power.

While the results of focusing efficiency measurements can be validated
through local deflection efficiency measurements, it is important to
note that deflection efficiency measurement cannot replace the
focusing efficiency measurement due to the loss of phase information
in the former approach~\cite{williams2023metalenses}. However, local
deflection efficiency measurements can provide additional insights and
complement the focusing efficiency data. Furthermore, the holographic
complex-field measurement technique introduced in this study has
valuable applications in troubleshooting and optimizing the design and
fabrication process. It offers direct information about the local
optical field, which cannot be obtained through intensity-based
characterization techniques alone. Therefore, this holographic
complex-field measurement technique can be employed for thorough
analysis, fine-tuning, and optimization of metasurface design and
fabrication methods.

The high focusing efficiency demonstrated here indicates the
feasibility of cascaded multilayer meta-optics using metasurfaces
fabricated through the NIL technique. In such systems, the total
efficiency is the product of the efficiency of each individual
metasurface element. Notably, the fabrication of separate metasurface
wafers through NIL allows for wafer-level alignment. This alignment
can be achieved using 3D metasurface alignment
marks~\cite{mcclung2023holographic}, which offer sub-micron
accuracy. As a result, fully-integrated optical metasystems can be
realized, eliminating the need for post-fabrication alignment and
assembly processes. This advancement enables the development of highly
efficient and compact optical systems that can be seamlessly
integrated into various applications.

\bibliographystyle{plain}

\vspace{0.2in}
\section*{Acknowledgements} 
This work was supported by the DARPA Extreme Optics and Imaging
program. Fabrication was performed at the core facilities of the
Institute for Applied Life Sciences (IALS) at the University of
Massachusetts Amherst and at the Center for Nanoscale Systems (CNS) at
Harvard University, which is part of the National Nanotechnology
Coordinated Infrastructure Network (NNCI). The NNCI is supported by
the National Science Foundation under NSF Award No. 1541959.

\setcounter{equation}{0}
\setcounter{figure}{0}
\newpage
\maketitle
\section*{Supporting Information}

\textbf{Master design.} We determined the design curve (Figure~2c) using a method described in a previous publication~\cite{arbabi2020increasing}. In the design, we assumed refractive indices of 2.02 for silicon nitride and 1.46 for fused silica. We explored different designs with varying nano-post heights and lattice constants. Using the grating averaging technique, we evaluated the focusing efficiency of metalenses with a numerical aperture (NA) of 0.2 for each design. Among the designs considered, we selected the one shown in Figure~2a, which exhibited a simulated focusing efficiency of 95\%.

\textbf{Master mold fabrication.} To make the master mold, we first deposited a 200-nm-thick amorphous silicon film on a 0.5-mm-thick fused silica substrate using a PECVD (PECVD, STS). We then spun on a positive tone electron beam resist (ZEP 520A, Zeon Specialty Materials) and a charge dissipation layer (ARPC 5090, Allresist) and exposed the metalens pattern using an electron beam lithography system (ELS-HS50, Elionix). We removed the charge dissipation layer, developed the resist, and transferred the pattern, which consisted of holes, to the amorphous silicon film by dry etching (ICP RIE, STS). We removed the resist using a solvent (Remover PG, Kayaku Advanced Materials) and treated the surface using an oxygen plasma asher (SCE-104, Anatech USA). We fluorinated the master mold by exposing it to heptadecafluoro-1,1,2,2-tetrahydrodecyltrichlorosilane (Gelest) in a desiccator pumped to rough vacuum and left overnight. Finally, we glued the master mold to a glass microscope slide using a UV-curable adhesive (NOA 84, Norland Products Inc.). Figure~3e shows an image of the master mold at this stage. 

\textbf{Stamp fabrication.}{
To cast a stamp, we first spun a layer of PDMS (Sylgard 184, Dow) on a flexible, 200-$\upmu$m-thick glass substrate (AF-32, Schott) to a thickness of ca. 140 $\upmu$m, and baked the sample overnight at 80 ºC. 
After the PDMS was cured, we cut a channel approximately the width of the metalens pattern into the PDMS layer using a razor blade. 
We then prepared an hPDMS solution consisting of 1.7 g of 7-8\% vinyl methysiloxane-dimethylsiloxane (Gelest), 5 $\upmu$L of 2,4,6,8-tetramethyl-2,4,6,8 tetravinylcyclotetrasiloxane (Sigma Aldrich), 9 $\upmu$L of platinum-divinyltetramethyldisiloxane (Gelest) and 5 g of toluene. 
We vortex mixed this solution, added 0.5 mL of 25-30\% methylhydrosiloxane-dimethylsiloxane (Gelest), and mixed again. We spun this solution onto the master mold at 3 kRPM for 40 seconds and baked the sample in an oven at 100 $^\circ$C for 45 minutes. We then poured a puddle of uncured PDMS (Sylgard 184, Dow), into the channel cut into the PDMS on glass workpiece, taking care to avoid forming bubbles, and placed the master mold, hPDMS side down, in this puddle, pressing to ensure the edges of the master mold were flush with the spun and cured PDMS. We baked this assembly overnight in an oven at 80 $^\circ$C. To release the stamp, we let the assembly cool, submerged it in a dish of isopropyl alcohol, and gently flexed the glass backing away from the master mold, holding the master mold against the bottom of the dish using the slide. We removed the stamp from the isopropyl alcohol, rinsed it with deionized water, and blew it dry with nitrogen. Figure~3f shows an image of a stamp at this stage.}

\textbf{Metalens fabrication.} To make the NIL metalenses, we first deposited a ca. 650-nm-thick silicon nitride film on a 0.5-mm-thick fused silica substrate using a PECVD (PECVD, STS). We used a spectroscopic ellipsometer (RC2, J.A. Woollam Ellipsometry Solutions) to measure the film’s thickness (667 nm) and its refractive index, which is shown in Figure~\ref{fig:refractive}. We then spun on a 90-nm-thick layer of PMMA A2 (Kayaku Advanced Materials) and soft baked the sample on a hotplate at 180 $^\circ$C for 3 minutes. We used a thermal imprinting tool (NX-2600BA, Nanonex) to transfer the pattern on the stamp to the PMMA layer. The imprint process consists of a 2 minute pump; a 50 second pre-imprint phase at 180 $^\circ$C and 100 psi; a 1 minute, 30 second imprint at 210 $^\circ$C and 200 psi; and a cooldown phase, after which the pressure is released. Pressures and temperatures during a typical imprint are shown in Figure~\ref{fig:thermal}. We carefully separated the stamp and workpiece after removing them from the tool. We then descummed the PMMA using oxygen plasma (Vision 320, Advanced Vacuum), deposited a 30-nm-thick aluminum oxide hard mask using electron beam evaporation, lifted off the PMMA in a solvent (Remover PG, Kayaku Advanced Materials) and sonicated the sample in acetone. Subsequently, we transferred the metalens pattern to the silicon nitride layer by dry etching in a mixture of \ce{SF_6} and \ce{C_4F_8} gases (PlasmaPro 100 Cobra, Oxford Instruments). Finally, we removed the aluminum oxide mask using a TMAH-based developer (AZ 300 MIF, AZ Electronic Materials). We fabricated the control metalens using an electron beam lithography process similar to the one for the master mold described above, using PECVD silicon nitride substrates instead of the amorphous silicon substrate. After developing the electron beam resist, we evaporated a 30-nm-thick aluminum oxide hard mask, lifted it off, and etched the control metalens similarly to the NIL metalenses.

\textbf{Focal spot characterization.} We obtained images of the focal spots using the setup shown schematically in Figure~4a. The light from a 552 nm laser (OBIS 552-20LS, Coherent) was passed through a 10$\times$ beam expander (Melles Griot, 09lbx003) and illuminated the metalenses. A custom-built microscope formed of a 50$\times$ objective (LMPlanFL N, Olympus) and 200 mm focal length tube lens (AC254-200-A-ML, Thorlabs) relayed a magnified image onto a CCD camera (CoolSnap K4, Photometrics). 

\textbf{Focusing and deflection efficiency measurements.}{
We used five lasers (532 nm: COMPASS 215M-15, Coherent; 552 nm: OBIS 552-20LS, Coherent; 559 nm: DPSS, Laserland Australia, 544 nm: HeNe 1676, JDS Uniphase; 633 nm: HeNe 1107P-2194, JDS Uniphase) in our focusing efficiency measurement. 
The laser beams were expanded using a 10$\times$ beam expander (Melles Griot, 09lbx003) and passed through a 6-mm-diameter aperture to match the metalens aperture. 
To measure the optical power focused by the metalens, we placed a 40-$\upmu$m-diameter pinhole (P40D, Thorlabs) in the focal plane of the metalens and measured the power passing through the pinhole using a power meter (PM100D with S120C head, Thorlabs). 
We corrected measured power for the percentage of the power on an ideal metalens blocked by the aperture (i.e., divided the measured value by 0.986). 
To determine the optical power incident on the metalens aperture, we remove the metalens and the pinhole, measured the power passing through the 6-mm-diamter aperture divide the result by 0.965 to correct for the reflection off the back side of the metalens substrate. 
Finally, the focusing efficiency was obtained by dividing the power of focused light by the incident power. 
To find the distribution of power inside the 40-$\upmu$m-diameter pinhole (Figure~\ref{fig:encircled}a), we constructed a high dynamic range composite image of the focal spot by combining many different exposures. We also measured the deflection efficiency (ratio of power in the first diffraction order to the incident power) of the metalens using the setup shown in Figure~\ref{fig:deflection}a: the light from a laser (559 nm, DPSS, Laserland Australia) was polarized and was focused on the plane of the metalens by passing through a 200 mm focal length lens (AC254-200-A-ML, Thorlabs). The loosely focused light ensured that the diffraction orders were well separated. The power of the first diffraction order and the incident beam were measured using a power meter (PM100D with S120C head, Thorlabs) and the incident beam power was divided by 0.965 to correct for the reflection off the back side of the metalens substrate. Finally, the deflection efficiency was obtained by diving the power of the first diffraction order by the incident beam power.}

\textbf{Holographic measurement.} {
We determined the phase of light at the metalens surface using the interferometric imaging setup shown schematically in Figure~5a. We illuminated the metalens with laser light (552 nm, OBIS 552-20LS, Coherent) and, using a custom-built microscope formed of a 10$\times$ objective (MPlanFL N, Olympus) and 200 mm focal length tube lens (AC254-200-A-ML, Thorlabs), relayed an image of its surface onto a transmissive diffraction grating with 300 grooves/mm (GT25-03, Thorlabs). A second image relay comprising a 75 mm focal length lens (LA1145, Thorlabs) and 500 mm focal length lens (LA1380, Thorlabs) projects this image onto a camera (CoolSnap K4, Photometrics). A 30-$\upmu$m-diameter pinhole (P30D, Thorlabs) in the Fourier plane of the relay acts as a low pass filter on the image from the first order, creating a clean reference beam for the interferogram. A screen blocks all other orders of the grating except for the zeroth order. We use a computational Fourier-domain filtering technique to retrieve the complex-valued image~\cite{bhaduri2014diffraction}. We used a freely available image processing library~\cite{van2014scikit} to unwrap the phase of the complex image. This unwrapped phase is shown in Figure 5b. The target phase for the metalens is given by $\phi(r) = k_0\left(f^2 + r^2\right)^{1/2} + \phi_0$, where $k_0$ is the free-space wavenumber, $f$ is the focal length of the metalens at its design wavelength, $r$ is the radial distance from the metalens center and $\phi_0$ is the phase offset at the center of the metalens. The data in Figures 5c and 5d show the difference of this target phase and the measured phase data.}

\textbf{Imaging.} We measured the spectra of the LEDs used as illumination sources in the imaging experiment (Figure~6b) using an optical spectrum analyzer (Ando AQ-6315a). The lens used in the reference image in Figure~6c was an aspheric lens with 5 mm clear aperture and an NA of 0.16 (C260TM, Thorlabs). All the images were captured on a CMOS camera with a color sensor (DCC1240C, Thorlabs). Figure~\ref{fig:camera} shows a photograph of the metalens on its camera mount.

\bibliographystyle{plain}

\newpage
\renewcommand{\thetable}{S\arabic{table}} 
\begin{table}
\centering 
\caption{Focusing efficiencies for the NIL and EBL metalenses. Errors represent uncertainty due to fluctuation in laser power.}
\begin{tabular}{ |c||c|c|  }
 \hline
Wavelength (nm) & Metalens & Focusing efficiency (\%)\\
 \hline
 \multirow{2}{*}{532} & NIL & 70$\pm$1 \\ 
& EBL & 77$\pm$1 \\ 
\hline 
 \multirow{2}{*}{543} & NIL & 73$\pm$1 \\ 
& EBL & 82$\pm$2 \\ 
\hline
\multirow{2}{*}{552} & NIL & 81$\pm$1 \\ 
& EBL & 89$\pm$1 \\ 
\hline
\multirow{2}{*}{559} & NIL & 80$\pm$2 \\ 
& EBL & 89$\pm$2 \\ 
\hline
 \multirow{2}{*}{633} & NIL & 68$\pm$1 \\ 
& EBL & 74$\pm$1 \\ 
\hline
\end{tabular}
\end{table}
\clearpage

\newpage
\renewcommand{\figurename}{Figure}
\renewcommand{\thefigure}{S\arabic{figure}} 

\begin{figure}
    \centering
    \includegraphics[width=\textwidth]{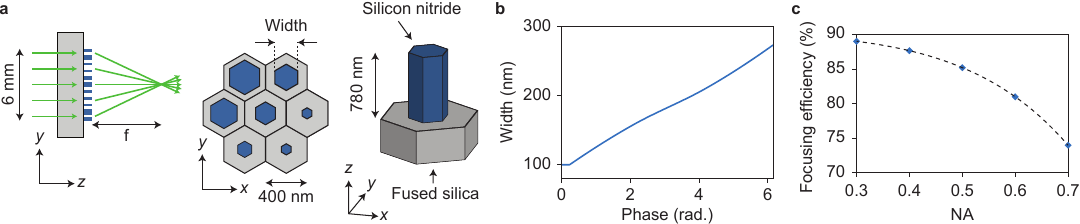}
    \caption{High NA silicon nitride metalenses with high focusing efficiency. (a) Schematic of the metalenses and the meta-atoms comprising the metalenses. (b)  The design curve that maps the desired phase to the nano-post width. The minimum nano-post width and the gap between the nano-posts are 100 nm and 120 nm, respectively. (c) Simulated focusing efficiency values for metalenses with different NAs that are designed using the meta-atoms shown in (a) and the design curve shown in (b). The simulation is performed using the grating averaging technique~\cite{arbabi2020increasing}. The dashed line is a fit to simulated data points (blue circles).} 
    \label{fig:high_NA_metalens}
\end{figure}

\clearpage

\begin{figure}
    \centering
    \includegraphics[width=\textwidth]{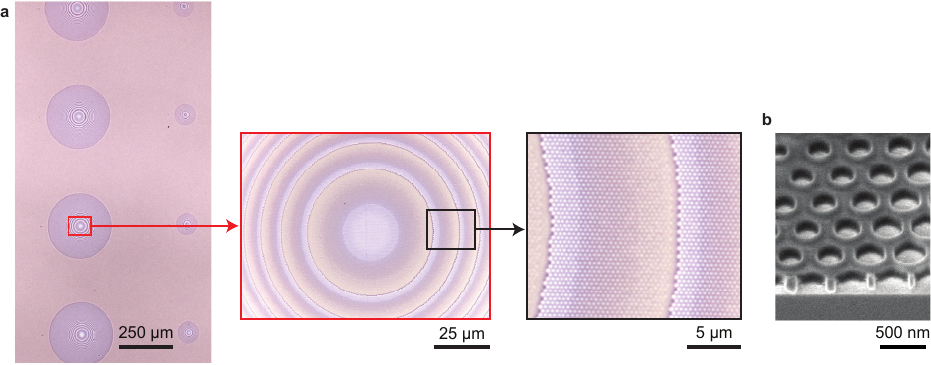}
    \caption{Images of imprinted PMMA. (a) Optical microscope, and (b) SEM images of imprinted PMMA.}
    \label{fig:pmma}
\end{figure}

\clearpage
\begin{figure}
    \centering
    \includegraphics[width=0.9\textwidth]{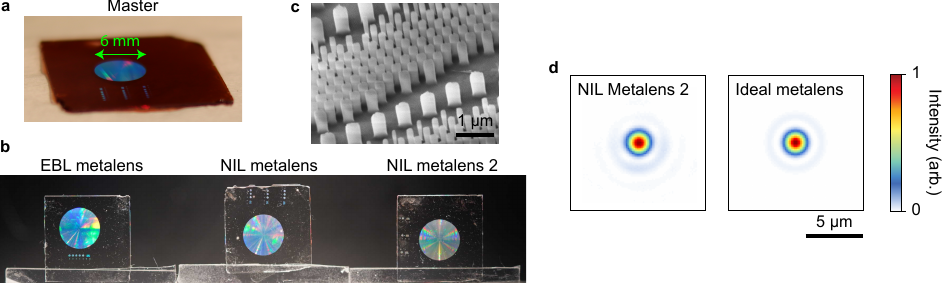}
    \caption{(a) A photo of the master mold used for fabricating the NIL metalenses. (b) Photos of the EBL metalens, the NIL metalens, and a second NIL metalens (NIL metalens 2). The NIL metalens 2 was fabricated prior to fully optimizing the silicon nitride etch process. (c) A scanning electron microscope image of the nano-posts on the NIL metalens 2 that shows a sub-optimal etch process. (d) Measured focal spot for the NIL metalens 2 and an ideal metalens. The negligible aberrations of the focal spot indicate that the stamp was not deformed significantly during the imprinting process. The measured focusing efficiency of the NIL metalens 2 was 73\%, indicating the effect of the sub-optimal etch. }
    \label{fig:second_NIL}
\end{figure}

\clearpage
\begin{figure}
    \centering
    \includegraphics{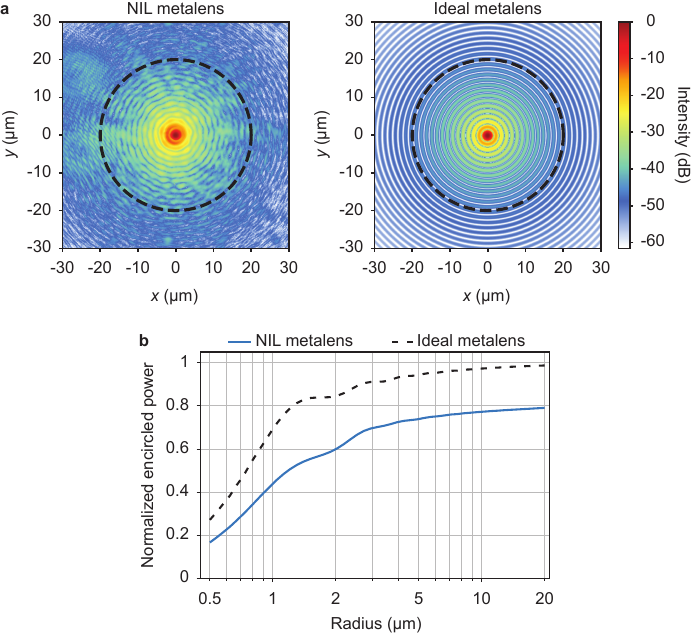}
    \caption{High dynamic range spot image. (a) Logarithmic scale plot of the measured focal spot at $\lambda$ = 559 nm for NIL metalens and the simulated focal spot for an ideal metalens with the same NA. Circle (black, dashed) shows the size of the 40-$\mu$m-diameter pinhole. (b) Relative encircled power for the NIL and ideal metalenses normalized to the power incident on the metalenses. }
    \label{fig:encircled}
\end{figure}

\clearpage
\begin{figure}
    \centering
    \includegraphics[width=\textwidth]{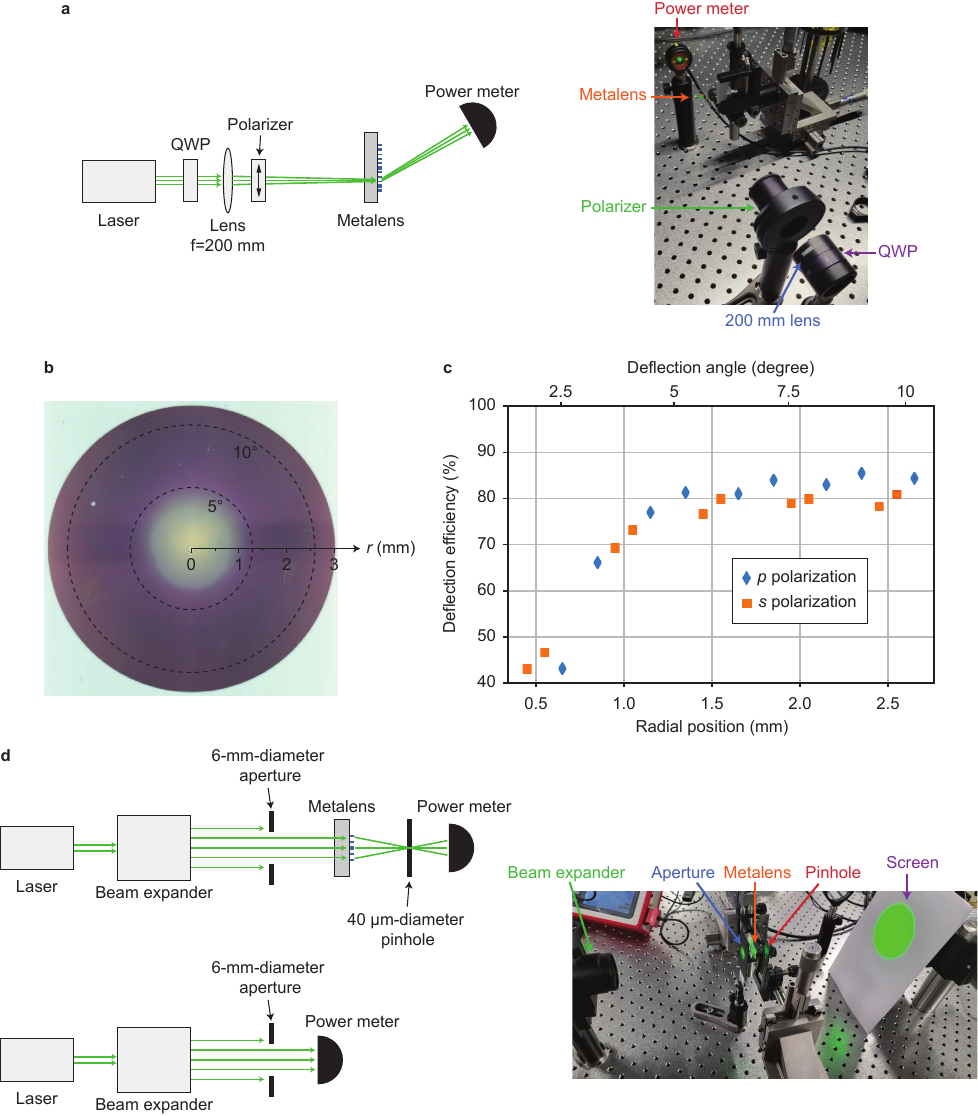}
    \caption{Deflection efficiency. (a) Schematic illustration and a photograph of the deflection efficiency measurement setup. (b) Optical microscope image of the NIL metalens used for deflection efficiency measurement. The radial distance and the locations on the metalens corresponding to deflection angles of 5$^\circ$ and 10$^\circ$ are marked. (c) Measured deflection efficiency for p and s polarized light. (d) Schematic illustration and a photograph of the focusing efficiency measurement setup. Efficiency values are presented in Figure 4e. QWP: Quarter-wave plate.}
    \label{fig:deflection}
\end{figure}

\clearpage
\begin{figure}
    \centering
    \includegraphics[width=\textwidth]{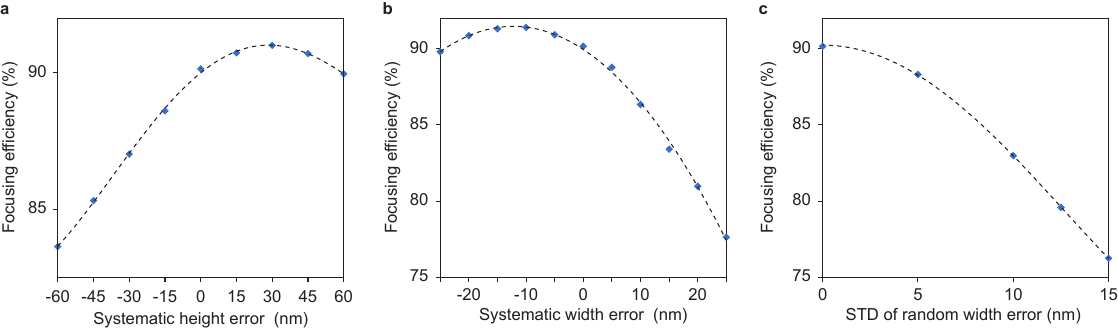}
    \caption{Sensitivity of metalens focusing efficiency to fabrication errors. (a) The  focusing efficiency of the metalens shown in Fig. 2a as a function of systematic (deterministic) errors in the height, and (b) width of all nano-posts. (c) The focusing efficiency of the metalens shown in Fig. 2a to random errors in nano-post widths. The error distribution is assumed to have a Gaussian distribution and the abscissa represents the standard deviation (STD) of the distribution. The simulations are performed using the grating averaging technique~\cite{arbabi2020increasing}. The dashed line are fits to simulated data points (blue circles).}
    \label{fig:sensitivity}
\end{figure}

\clearpage
\begin{figure}
    \centering
    \includegraphics{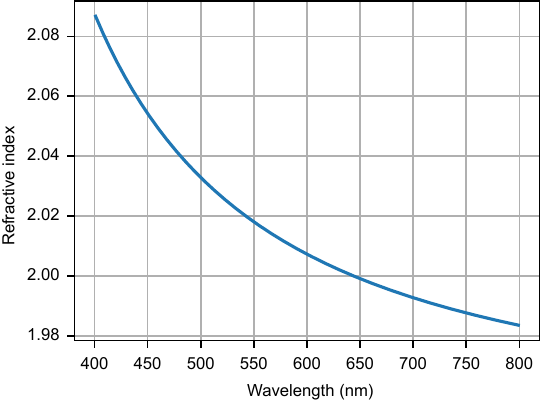}
    \caption{Refractive index of silicon nitride. The refractive index values are measured using a spectroscopic ellipsometer.}
    \label{fig:refractive}
\end{figure}

\clearpage
\begin{figure}
    \centering
    \includegraphics[width=\textwidth]{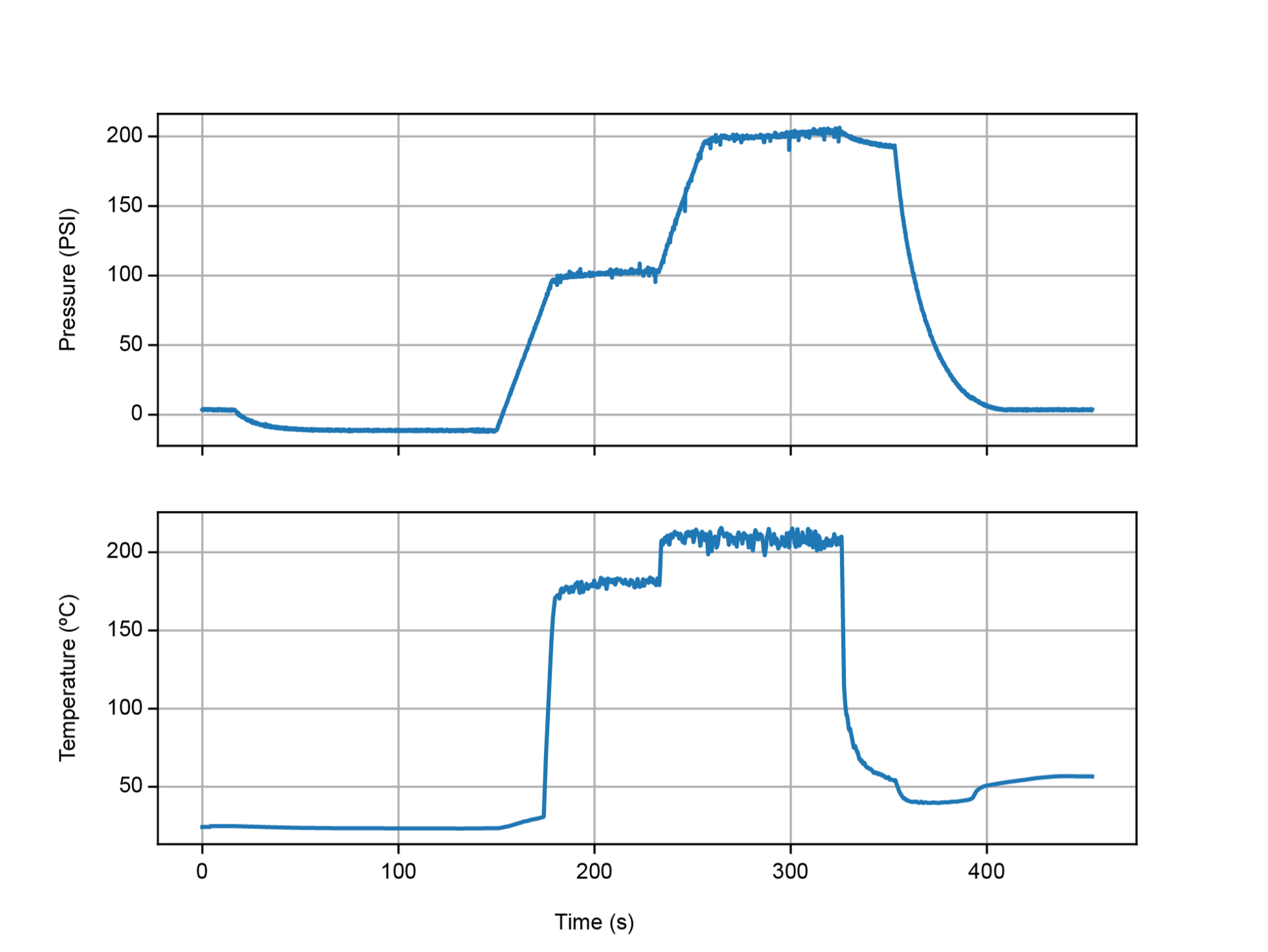}
    \caption{Thermal imprint. Pressure (top) and temperature (bottom) as a function of time.}
    \label{fig:thermal}
\end{figure}

\clearpage
\begin{figure}
    \centering
    \includegraphics[width=0.6\textwidth]{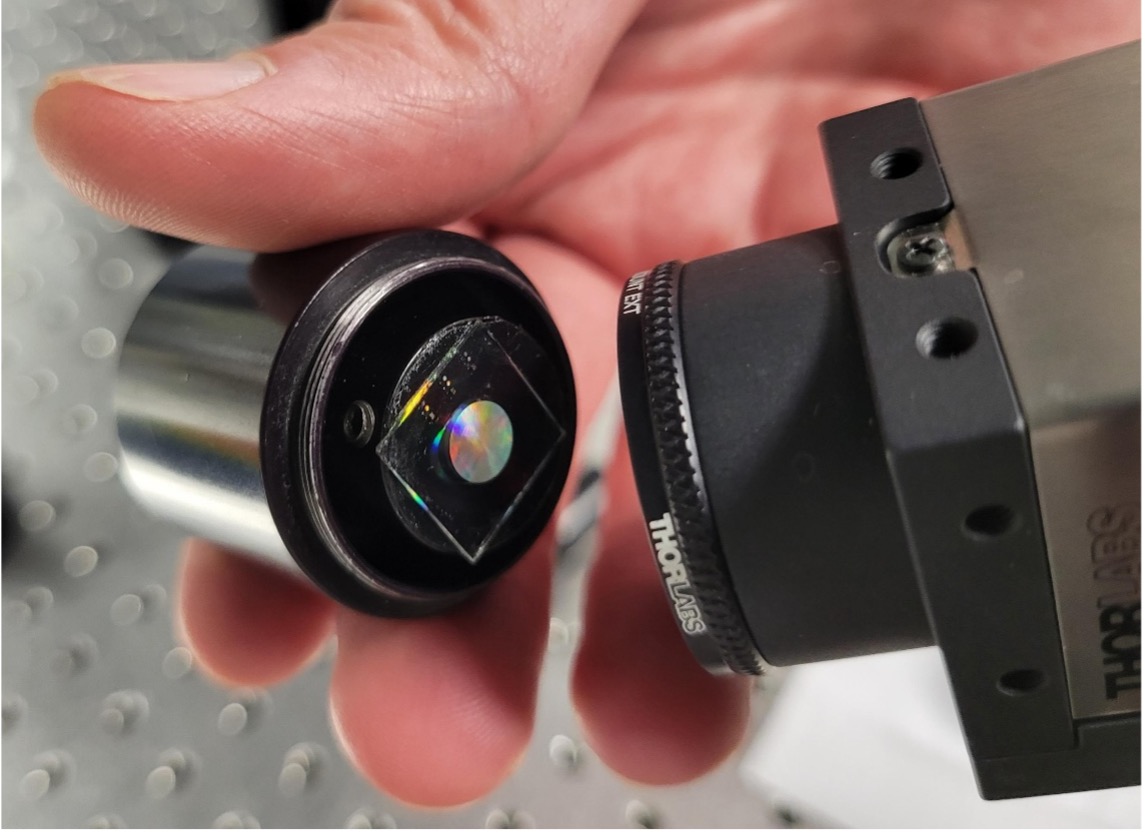}
    \caption{Photograph of metalens camera.}
    \label{fig:camera}
\end{figure}

\end{document}